\documentclass[onecolumn,authoryear]{els-mrw} 

\usepackage{amsmath,amssymb,amsfonts,amsthm,makeidx,graphicx}
\usepackage{txfonts}
\usepackage{helvet}
\usepackage{comment}
\usepackage{hyperref}

\begin{document}

\chapter{Mergers of galaxies}\label{chap1}

\author[1]{Sugata Kaviraj}

\address[1]{\orgname{Centre for Astrophysics Research}, \orgdiv{University of Hertfordshire}, \orgaddress{College Lane, Hatfield AL10 9AB, UK}}


\maketitle

\textbf{Keywords:} Galaxy mergers; Star formation; Active Galactic Nuclei, Galaxy morphological classification, Early-type galaxies.


\begin{glossary}[Glossary]
\begin{tabular}{@{}lp{34pc}@{}}
\term{Active galactic nucleus} & A black hole that has been triggered by the accretion of matter and is emitting electromagnetic radiation.\\
\term{Dark matter halo} & A gravitationally bound region of dark matter.\\
\term{Cosmic variance} & The uncertainty in the measured number density of galaxies or other extragalactic objects due to large-scale density fluctuations in the Universe.\\
\term{Cosmological simulation} & A large-volume simulation of the Universe which can produce statistical samples of mock galaxies that can be compared to large populations of galaxies in observational surveys.\\
\term{Intergalactic medium} & Matter that occupies the space between galaxies.\\
\term{Interstellar medium} & Matter that occupies the space between stars in a galaxy.\\
\term{Major merger} & A merger between two galaxies where the stellar mass ratio of the merging galaxies is greater than 1:4 (i.e. the stellar masses of the two galaxies are relatively similar).\\
\term{Massive galaxy} & Defined here as a galaxy that has a stellar mass ($M_{\star}$) greater than $10^{10}$ M$_{\odot}$.\\
\term{Minor merger} & {\color{black}Defined here} as a merger between two galaxies where the stellar mass ratio of the merging galaxies is less than 1:4 (i.e. the stellar masses of the two galaxies are quite dissimilar). Mergers with mass ratios less than around 1:10 typically only affect galaxies weakly, so often the term minor merger refers to mergers with mass ratios between 1:4 and 1:10.\\
\term{Stellar mass function} & The measured number of galaxies per unit volume as a function of stellar mass.\\
\term{Zoom-in simulation} & A re-simulation at very high resolution of an object selected from a lower resolution simulation.\\
\end{tabular}
\end{glossary}

\begin{glossary}[Nomenclature]
\begin{tabular}{@{}lp{34pc}@{}}
\term{$\Lambda$CDM} & $\Lambda$ cold dark matter.\\
\term{H} & Hydrogen.\\
\term{He} & Helium.\\
\term{Metals} & All elements in the periodic table other than hydrogen and helium.\\
\term{$M_{\rm \star}$} & The stellar mass of a galaxy.\\
\term{SFR} & The star formation rate of a galaxy i.e. the stellar mass formed per unit time.\\
\end{tabular}
\end{glossary}


\begin{abstract}[Abstract]
The evolution of our Universe is strongly influenced by the attractive force of gravity. A key aspect of this evolution, therefore, is the merging of galaxies. Here, we explore the role of mergers in shaping the properties of massive galaxies over cosmic time. Observational methods of finding mergers include identifying galaxy pairs in close proximity, visual inspection of galaxy images to identify signatures of mergers (e.g. tidal features) and using morphological parameters such as Asymmetry and the Gini coefficient. The fraction of merging galaxies increases with redshift, potentially out to $z\sim6$. The principal impact of merging is to transform the morphological mix of massive galaxies, from largely rotationally-supported systems at high redshift to more dispersion-dominated systems in the nearby Universe. Mergers also drive gas towards the central regions of the remnant, fuelling starbursts and feeding supermassive black holes. However, only around a third of the stellar and black hole mass at the present day is directly attributable to merging. 
\end{abstract}

\section*{Learning objectives}

\begin{itemize}
    \item Understand why mergers are important in the evolution of galaxies, in the context of the standard paradigm of hierarchical structure formation. 
    \item Understand how mergers are identified, both in observations and in numerical simulations of galaxy formation.
    \item Gain an overview of how mergers drive the morphological transformation of massive galaxies over cosmic time.
    \item Gain an overview of how mergers impact star formation and black-hole growth in massive galaxies over the lifetime of the Universe.
\end{itemize}


\section{Introduction - the hierarchical formation of galaxies over cosmic time}
\label{sec:theory}
In the currently-accepted $\Lambda$ cold dark matter ($\Lambda$CDM) paradigm of structure formation, primordial density perturbations, that are seeded by quantum fluctuations and amplified by inflation \citep[e.g.][]{Guth1981}, grow under the influence of gravity to form dark matter halos \citep[e.g.][]{Rees1977,White1978}. Smaller halos form first and merge under the influence of gravity to form progressively larger ones \citep[e.g.][]{Peebles1982,Blumenthal1984}. Baryonic matter (initially pristine hydrogen and helium) condenses on to the dark matter halos and settles into rotationally-supported disc-like distributions as it cools and contracts. Cold gas fuels star formation \citep[e.g.][]{Gao2004,Bigiel2008}, with the star formation rate (SFR) being governed by the local gas density \citep[e.g.][]{Schmidt1959,Kennicutt1998}. As stellar assembly progresses, supernovae enrich the interstellar medium with metals, and inject thermal and kinetic energy into the ambient gas \citep[e.g.][]{Scannapieco2008,Cen1999}. In addition, the infall of material on to central supermassive black holes releases part of the rest-mass energy of the accreted matter into the gas reservoir of their host galaxies \citep[e.g.][]{Fabian2012,Beckmann2017}. The energetic feedback from supernovae and the supermassive black hole acts to regulate star formation activity and the stellar mass growth of the galaxy. The evolution of a galaxy, over cosmic time, is driven by two principal processes: accretion from the intergalactic medium and mergers and interactions with other galaxies. 

Mergers are thought to impact galaxies in three principal ways. They influence stellar mass growth, either by directly adding stars formed ex-situ to the galaxy and, in some cases, by enhancing the SFR, which accelerates stellar mass growth beyond what would have been possible had the galaxy not been merging. Mergers may also increase the accretion rate of the central supermassive black hole, accelerating its growth. However, the most important impact of mergers is in facilitating morphological transformation. Mergers make stellar orbits more chaotic, which has the effect of `washing out' pre-merger structure (e.g. spiral arms) that may have been present in the merging progenitors.  

However, the extent to which each of these changes takes place depends on the properties of the merger itself, e.g. the stellar mass ratio of the merging galaxies, their orbital geometry and the gas fractions of the merging systems \citep[e.g.][]{Martin2021}. For example, `major' mergers, defined here as having stellar mass ratios greater than 1:4, typically produce more significant morphological changes and larger enhancements in star formation and black-hole accretion rates than `minor' mergers (defined here as having stellar mass ratios less than 1:4). Similarly, mergers that involve larger gas fractions are likely to produce more significant increases in star formation activity and black-hole growth. 

It is worth noting that a natural consequence of hierarchical structure formation is that galaxies with lower stellar masses outnumber their more massive counterparts. For example, the stellar mass function at $z\sim0$ indicates that the number density of galaxies with stellar masses around $10^8$ M$_{\odot}$ is more than an order of magnitude larger than that of objects with stellar masses around $10^{11}$ M$_{\odot}$ \citep[e.g.][]{Weaver2023}. As a result, major mergers are rarer than minor mergers at all epochs and in all environments. Thus, while individual major mergers are more spectacular and capable of producing more significant changes in the merger remnant, the greater frequency of minor mergers means that, over cosmic time, they are capable of influencing the galaxy population to a similar degree as their major counterparts. 

This chapter begins by describing how galaxy mergers are identified in cosmological simulations and observational data, followed by a discussion of the incidence of mergers over cosmic time. It then describes the impact of mergers on three key aspects of galaxy evolution - morphological transformation, star formation and black hole growth. Our focus will be on data from large-scale observational surveys and simulations with cosmological volumes. As such, the discussion relates largely to the broader galaxy population which lives in relatively low-density environments, and not dense regions, like clusters, which contain a minority of galaxies in the Universe. 

Finally, the narrative will focus on massive ($M_{\star} \gtrsim 10^{10}$ M$_{\odot}$) galaxies because they are bright enough to be detectable in past surveys across a large fraction of cosmic time. While low-mass or `dwarf' ($M_{\star} \lesssim 10^{10}$ M$_{\odot}$) galaxies dominate the galaxy number density, typical dwarfs are undetectable outside the local neighbourhood in previous surveys like the SDSS \citep[][]{Alam2015}, which have large footprints but are shallow. The small subset of dwarfs that appear in these surveys outside the very local Universe have high SFRs, which enhance their luminosities, making them detectable in shallow surveys \citep[e.g.][]{Jackson2021b}. But the bias towards star-forming systems also makes it difficult to use these subsets to understand the evolution of the dwarf population as a whole \citep[e.g.][]{Kaviraj2025}. As a result, our current statistical understanding of galaxy evolution is overwhelmingly dominated by massive galaxies, which also form the basis of the discussion in this chapter.


\section{Identification of mergers}


\subsection{Mergers in simulations}

\begin{figure}
  \centering
  \includegraphics[width=0.6\textwidth]{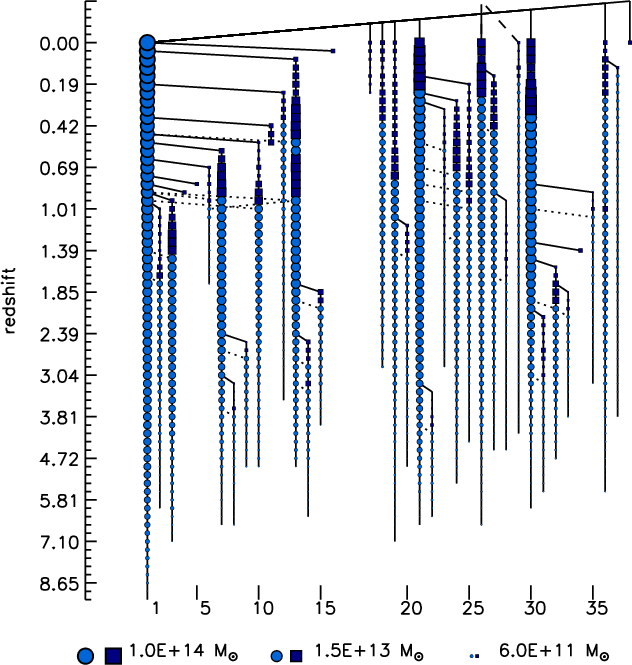}
  \caption{Example of a merger tree from \citet{Tweed2009} of a dark matter halo which has a mass of $10^{14} \rm M_{\odot}$ at $z=0$. Circles represent halos while squares show `subhalos' (i.e. halos that are already embedded in larger halos but have not been completely disrupted yet). The merger tree of the main object in question is shown on the left hand side, while the merger trees of objects that it has accreted are shown via the branches to the right. Mergers between halos and their subhalos are indicated using solid lines and mergers between subhalos are indicated by dotted lines. The sizes of the symbols scale with their masses (see legend at the bottom of the figure). Only the 40 most massive branches of the full merger tree are shown for clarity. Credit: Figure 9 from Dylan Tweed et al., A\&A, 506, 647-660, 2009, reproduced with permission $\copyright$ ESO.}
  \label{fig:merger_tree}
\end{figure}

In simulations, dark matter halos and their constituent galaxies at consequent time steps can be linked, producing a record of the merger history of a galaxy, which is often visualised via `merger trees'. Figure \ref{fig:merger_tree} shows an example of such a merger tree of a dark matter halo which has a mass of $10^{14} \rm M_{\odot}$ at $z=0$ \citep{Tweed2009}. Circles represent halos while squares show `subhalos' (halos that are already embedded in larger halos but have not been completely disrupted yet). The merger tree of the main object is shown on the left hand side, while the merger trees of systems that it has accreted are shown via the branches to the right. Mergers between halos and their subhalos are shown using solid lines and mergers between subhalos are indicated using the dotted lines. Only the 40 most massive branches of the full merger tree are shown in this figure. Figure \ref{fig:merger_tree} demonstrates the richness of the merger history experienced by a typical massive dark matter halo (and its constituent galaxy). As we shall see in Section \ref{sec:incidence_mergers}, while construction of merger trees is relatively straightforward for mock galaxies in simulations, comparison of the incidence of mergers to observational estimates can nevertheless be a non-trivial exercise. 

\subsection{Mergers in observations}

While the merger history and the impact of merging in simulated galaxies can be studied to arbitrary precision (within the resolution limits of the simulation in question), the empirical study of galaxy mergers is more challenging, both in terms of securely identifying galaxies that are experiencing mergers and quantifying the impact of those mergers on the galaxy. Observationally, mergers have to be identified either through their structural impact on the system, e.g. via internal asymmetries or tidal features induced by the merger, or the fact that a galaxy is in very close proximity to another object, even if no structural distortions are visible. It is worth noting here that, notwithstanding the richness of the merger histories of massive galaxies (as shown in Figure \ref{fig:merger_tree}), the observational signatures of merging are dominated by the most recent merger event. Hence, the changes observed in individual galaxies are mostly related to this event and not the historical events that have taken place over its lifetime. This issue can be circumvented by studying statistical populations of mergers at different redshifts over cosmic time, with the underlying assumption that galaxy populations at higher redshift are progenitors of their counterparts at lower redshift. In this section we outline some of the most common methods of merger identification using observational data. 


\subsubsection{Identification via close pairs}

A methodologically simple method of identifying galaxies that are potentially in the process of merging is to select pairs of galaxies that are in close proximity to each other, both in terms of their projected physical distances on the sky and the difference in their recessional velocities, measured via their redshifts \citep[e.g.][]{Carlberg1994,Patton2002,Lefevre2000,Lopezsanjuan2012,Tasca2014}. The threshold values of spatial and redshift proximity that are used to identify mergers are estimated using simulations. For example, around 90 per cent of pairs with small physical separations (less than around 20 -- 30 kpc) and low velocity differences (less than around 500 km s$^{-1}$) are likely to merge on timescales of around a Gyr \citep[e.g.][]{Kitzbichler2008,Patton2024}. Such values are commonly used to identify merging systems via close pairs in observational work. 




\subsubsection{Identification via visual inspection - tidal bridges, tails, streams and shells} 
\label{sec:visual_inspection}

Visual inspection has been commonly used for classifying the massive galaxy population into its principal morphological classes. While we consider galaxy morphology in the context of mergers in more detail in section \ref{sec:morphological_transformation} below, we introduce some key concepts here, since they are relevant to the discussion in this section. The morphological mix of massive galaxies -- originally visualised in Hubble's `tuning fork' diagram \citep{Hubble1936} and consolidated in more recent work \citep[e.g.][]{Sandage1961} -- is composed of two main visually-identified morphological classes: `late-type' and `early-type' galaxies. While late-types are largely dominated by rotation, early-types are more dispersion-dominated. Structurally, late-types typically show features like spiral arms that are created via rotation, while early-types show a dominant bulge component (typified by a central light concentration) and an otherwise smooth light distribution. As we shall see later, the process of merging provides a link between these morphological classes. 
While visual inspection has been an important method of classifying galaxies into late and early-types, it can also be used to identify galaxies that are merging (without the need for positional or redshift information, as is the case for close pairs). 

\begin{figure}
  \centering
  \includegraphics[width=0.8\textwidth]{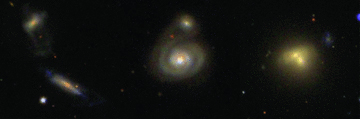}
  \caption{Example images of pre-mergers, from \citet{Darg2010}, at different stages of approach, from the SDSS. Standard depth SDSS imaging has a surface-brightness limit of around 26.4 mag arcsec$^{-2}$ \citep[e.g.][]{Kniazev2004}. Note the presence of tidal bridges connecting the merging galaxies in the left and middle panels. Credit: Figure 2 in Daniel W. Darg et al. `Galaxy Zoo: the properties of merging galaxies in the nearby Universe - local environments, colours, masses, star formation rates and AGN activity.' Monthly Notices of the Royal Astronomical Society (401) 3 (2010): 1552–1563.}
  \label{fig:approaching_mergers}
\end{figure}

Visual inspection typically identifies merging systems at two main stages. The first are pre-mergers -- these are systems in which the two merging galaxies are interacting but have not yet coalesced. Such systems are identifiable through the presence of `tidal bridges' -- streams of stars connecting the merging galaxies that have been ejected by the tidal forces created by the merger. Figure \ref{fig:approaching_mergers} presents examples of pre-mergers, with the two merging galaxies at various stages in their approach towards each other. The second are post-mergers -- these are systems where the merger is complete but the recently created merger remnant is identifiable via tidal structures (e.g. tidal tails, streams and shells) that have been created as a result of the merger. 

Tails are created by material ejected from the more massive (`primary') companion and therefore share the properties of the stars that reside in the outskirts of this primary galaxy. Streams, on the other hand, are produced due to the disruption of the less massive (`secondary') companion and therefore do not share the characteristics of the primary galaxy. While more equal mass mergers largely produce tidal tails, unequal mass ratio mergers result in a preponderance of streams around the remnant. However, both tidal tails and streams signpost the presence of a recent merger in equal measure. While orbits with low eccentricity tend to favour the production of tidal tails and streams \citep[e.g.][]{Amorisco2015}, shells are created by mergers which have relatively large mass ratios ($>$ 1:10), where the secondary companion is on a nearly-radial, low angular momentum trajectory \citep[e.g.][]{Pop2018}. Stars that are tidally stripped from the secondary oscillate in the gravitational potential of the remnant and, over time, wrap in phase space and accumulate at the apocentres of their orbits. These accumulations define sharp edges which are seen as shells {\color{black}which can have lifetimes of several Gyrs} \citep[e.g.][] {Quinn1984}. 

\begin{figure}
  \centering
  \includegraphics[width=\textwidth]{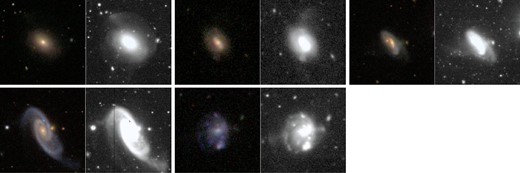}
  \caption{Examples of standard-depth, multicolour images from the SDSS (left) and their deeper $r$-band SDSS Stripe 82 counterparts (right), from \citet{Kaviraj2014b}, of merger remnants that have different morphologies -- early-type (row 1, left), Sa (row 1, middle), Sb (row 1, right), Sc (row 2, left) and Sd (row 2, middle) galaxies. The Sa, Sb, Sc and Sd classifications denote sub-classes of late-type galaxies in which the bulge components becomes increasingly less dominant. The massive galaxies in this study are at $z<0.07$. The Stripe 82 imaging is around 2 mag deeper than the standard-depth SDSS images. All galaxies in this figure exhibit extended tidal features. The galaxies in the left panel of row 1 and the right-hand panel of row 2 exhibit shell-like structures. The tidal features are virtually invisible in the shallower images, demonstrating the need for deep images for such an exercise. Credit: Figure 1 in Sugata Kaviraj `The importance of minor-merger-driven star formation and black hole growth in disc galaxies.' Monthly Notices of the Royal Astronomical Society (440) 4 (2014): 2944–2952.}
  \label{fig:postmergers1}
\end{figure}

\begin{figure}
  \centering
  \includegraphics[width=\textwidth]{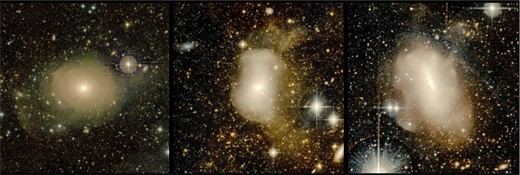}
  \caption{Examples of early-type galaxies which show evidence of post-merger signatures e.g. strongly perturbed shapes, tidal tails and shells, from \citet{Duc2015}. The galaxies shown are, from left to right, NGC 5557, NGC 1222 and NGC 2764. The galaxies in this study are at distances less than 42 Mpc. These images are around 3 mag deeper than standard-depth SDSS imaging. Credit: Figure 19 in Pierre-Alain Duc et al. `The ATLAS$^{3D}$ project – XXIX. The new look of early-type galaxies and surrounding fields disclosed by extremely deep optical images.' Monthly Notices of the Royal Astronomical Society (446) 1 (2015): 120-143. Reproduced with permission from the Oxford University Press on behalf of the Royal Astronomical Society.}
  \label{fig:postmergers2}
\end{figure}

Figure \ref{fig:postmergers1} shows deep SDSS Stripe 82 images of post-merger systems that have different morphologies ranging from early-type to various sub-classes of late-types, with all galaxies showing extended tidal features (see the caption of the image for more details). Figure \ref{fig:postmergers2} shows extremely deep images of post-merger early-type galaxies from the MATLAS survey \citep{Duc2015}, which exhibit strongly perturbed shapes, tidal tails and shells. It is worth noting that identifying mergers via tidal structures benefits from deep imaging, since these features tend to be faint and contribute only a few per cent of the total luminosity of the system \citep[e.g.][]{Martin2022}. Figures \ref{fig:approaching_mergers}, \ref{fig:postmergers1} and \ref{fig:postmergers2} present progressively deeper images, which demonstrate that tidal structures become easier to see as the image depth increases.

\begin{BoxTypeA}[chap1:box1]{Definition of common morphological parameters}

\noindent Asymmetry ($A$) parameter, as described in \citet{Abraham1996} and \citet{Conselice2003}, is defined as

\begin{equation}
    \rm \mathit{A}=\frac{\sum_{\mathit{i},\mathit{j}} \mid \mathit{I}_{ij} - \mathit{I}_{ij}^{180} \mid }{\sum_{\mathit{i},\mathit{j}} \mid \mathit{I}_{ij} \mid  } - \mathit{A}_{bgr},
    \label{eq:asy}
\end{equation}

\noindent where \textit{I$\rm_{i,j}$} and \textit{I$\rm_{i,j}^{180}$} are the pixel values of the original and the rotated images, respectively. $A$ is typically calculated within a circular aperture of either 1 or 1.5 $\times$ \textit{R}$\rm_{petro}$, which denotes the Petrosian radius \citep{Petrosian1976}. \textit{A}$\rm_{bgr}$ is the asymmetry of the background.\\

\noindent Clumpiness, as described in \citet{Lotz2004}, is defined as

\begin{equation}
    \rm \mathit{S}=\frac{\sum_{\mathit{i},\mathit{j}}  \mathit{I}_{ij} - \mathit{I}_{ij}^{S} }{\sum_{\mathit{i},\mathit{j}} \mathit{I}_{ij}} - \mathit{S}_{bgr},
    \label{eq:s}
\end{equation}

\noindent where \textit{I}$\rm_{i,j}$ and \textit{I}$\rm_{i,j}^{S}$ are the pixel values of the original image and its smoothed version, respectively, typically within circular apertures of 1 or 1.5 $\times$ \textit{R}$\rm_{petro}$. The smoothed image is obtained using a boxcar filter of width $\sigma$, which is set to 0.25 $\times$ \textit{R}$\rm_{petro}$ and S$\rm_{bgr}$ is the clumpiness of the background.\\ 

\noindent The Gini coefficient ($G$), as described in \citet{Lotz2004} is defined as 

\begin{equation}
    \rm \mathit{G}=\frac{1}{\mid\mathit{\overline{X}}\mid \mathit{n} (\mathit{n}-1)} \sum_{\mathit{i}=1}^{\mathit{n}} (2\mathit{i}-\mathit{n}-1) \mid \mathit{X}_i \mid
    \label{eq:gini}
\end{equation}

\noindent where \textit{n} is the number of pixels in the image, $\overline{X}$ is the mean of the pixel values, \textit{X}$\rm_i$ corresponds to the value of each pixel {\color{black}when ordered by increasing values of flux}.\\ 

\noindent The \textit{M}$_{20}$ index is the second order moment of the brightest 20 per cent of the galaxy's flux, divided by the total second order central moment ($\rm \textit{M}_{tot}$). As described in \citet{Lotz2004}, it is defined as

\begin{equation}
    \rm \mathit{M}_{tot}=\sum_\mathit{i}^\mathit{n} \mathit{M}_i=\sum_\mathit{i}^\mathit{n} \mathit{f}_i ((\mathit{x}_i-\mathit{x}_c)^2+(\mathit{y}_i-\mathit{y}_c)^2)
    \label{eq:m20}
\end{equation}

\noindent where \textit{f}$\rm_i$ is the flux of each pixel, \textit{x}$\rm_i$ and \textit{y}$\rm_i$ represent the positions of each pixel and \textit{x}$\rm_c$ and \textit{y}$\rm_c$ correspond to the location of the galaxy's centre, which are estimated by minimizing \textit{M}$\rm_{tot}$.\\

\end{BoxTypeA}


\subsubsection{Identification via morphological parameters} 

Another important method of identifying mergers is via morphological parameters that describe the principal structures in a galaxy image, without assuming a profile for the light distribution \citep[e.g.][]{Conselice2014}. The parameters are typically calibrated against the morphological classes derived via visual inspection. Using quantitative parameters automates the process of merger identification, making it better suited to the processing of large survey datasets. The identification of mergers has traditionally been underpinned by four parameters (see the boxed text for their mathematical definitions). Asymmetry ($A$; equation \ref{eq:asy}) is measured by rotating the galaxy image by 180 degrees and subtracting it from its original counterpart. Clumpiness ($S$; equation \ref{eq:s}) is calculated by subtracting a smoothed image from the original. The Gini coefficient ($G$; equation \ref{eq:gini}) measures the inequality in the distribution of light within a galaxy, while $M_{20}$ (equation \ref{eq:m20}) is the second order moment of the highest 20 per cent of a galaxy's flux. Combinations of these parameters are able to detect the changes (e.g. increases in asymmetry) that are induced by galaxy mergers. For massive galaxies, combining such morphological parameters via the relations  

\begin{equation}
(A > 0.35) \:\: \& \:\: (A > S)
\label{eq:as_merger}
\end{equation}

\begin{equation}
G > -0.14 \times M_{\rm 20} + 0.33
\label{eq:gm20_merger}
\end{equation}

enables us to identify systems that are experiencing significant mergers (which are typically highly star-forming galaxies in the nearby Universe), both at the first pass during the merger process and around the point where the merging galaxies are close to final coalescence \citep[e.g.][]{Conselice2006b,Lotz2008,Lotz2010}. 
While equation \ref{eq:as_merger} is most sensitive to major mergers, equation \ref{eq:gm20_merger} is able to probe lower mass ratios, down to around 1:9. {\color{black}The contamination from non-merging systems is typically low (at least in the nearby Universe) and these relations are able to identify at least 50 per cent of systems that are undergoing merging \citep[e.g.][]{Conselice2014}.} 


\section{The incidence of mergers over cosmic time}
\label{sec:incidence_mergers}

A fundamental metric that is used to measure merging activity is the merger fraction, which is defined as

\begin{equation}
f_{\rm merg} (M_*,z) = \frac{N_{\rm merg}}{N_{\rm total}},
\end{equation}

where $f_{\rm merg} (M_*,z)$ is the merger fraction\footnote{{\color{black}Note that, since the brightness of a galaxy is correlated with its stellar mass, the range of merger mass ratios that are visible will show some variation with $M_*$ and $z$. This is because the detection limit of the survey in question will lead to a lower limit for the stellar mass of galaxies that can be observed as a function of $z$.}} in the stellar mass and redshift ranges of interest and $N_{\rm merg}$ and $N_{\rm total}$ are the number of mergers and the total number of galaxies in these ranges. A rich literature has probed the redshift-evolution of $f_{\rm merg} (M_*,z)$, which is often parametrised in power law form as

\begin{equation}
f_{\rm merg} (z)= f_{\rm merg,0} \times (1+z)^m,
\label{eq:merger_fraction}
\end{equation}

where $f_{\rm merg,0}$ is the merger fraction at $z\sim0$. A related quantity is the merger rate which is defined as,

\begin{equation}
R_{\rm merg}= \frac{\phi_{\rm merg}}{T_{\rm merg}},
\end{equation}

where $\phi_{\rm merg}$ is the number density of galaxies identified as being part of a merger and $T_{\rm merg}$ is the timescale over which the merger can be observed. While $f_{\rm merg}$ can be calculated directly from observational data, $T_{\rm merg}$ cannot, because we observe only a snapshot of each individual merging system at some point during the merger process. Estimates of $T_{\rm merg}$ therefore have to be calculated using theoretical simulations, which adds an additional (and potentially significant) source of uncertainty to the derived value of $R_{\rm merg}$. Given the fact that $f_{\rm merg}$ is the more easily accessible observational quantity, we focus our subsequent discussion on this parameter. {\color{black}Readers are directed to \citet{Conselice2014} for more detailed discussions of $R_{\rm merg}$.} 

It is worth noting first that the measured value of $f_{\rm merg}$ can depend on factors like the depth of the survey, the redshift range considered, the adopted value of $f_{\rm merg,0}$, cosmic variance and the way galaxies are selected in the first place (e.g. using stellar mass vs luminosity). When survey characteristics are comparable and galaxy selection is performed in a similar way, the best-fit values of $m$ (in equation \ref{eq:merger_fraction}) for major mergers in massive galaxies, derived using the various identification methods described above, are greater than 1 and typically in the range $2<m<4$ \citep[see e.g.][]{Lefevre2000,Patton2002,Conselice2003,Kartaltepe2007,Bluck2009,Lotz2011}. In other words, in massive galaxies, the major merger fraction increases towards higher redshifts. The `zero-point' of this relation ($f_{\rm merg,0}$) is typically less than 10 per cent \citep[e.g.][]{Lopezsanjuan2012}. Integrating the merger activity over cosmic time, {\color{black}where measurements of $f_{\rm merg}$ are combined with values of $T_{\rm merg}$ from simulations as described above}, suggests that a typical massive galaxy will experience around one major merger in the redshift range $0<z<3$ \citep[which represents more than 80 per cent of the lifetime of the Universe, e.g.][]{Man2016,Conselice2022}. 

\begin{figure}
  \centering
  \includegraphics[width=\textwidth]{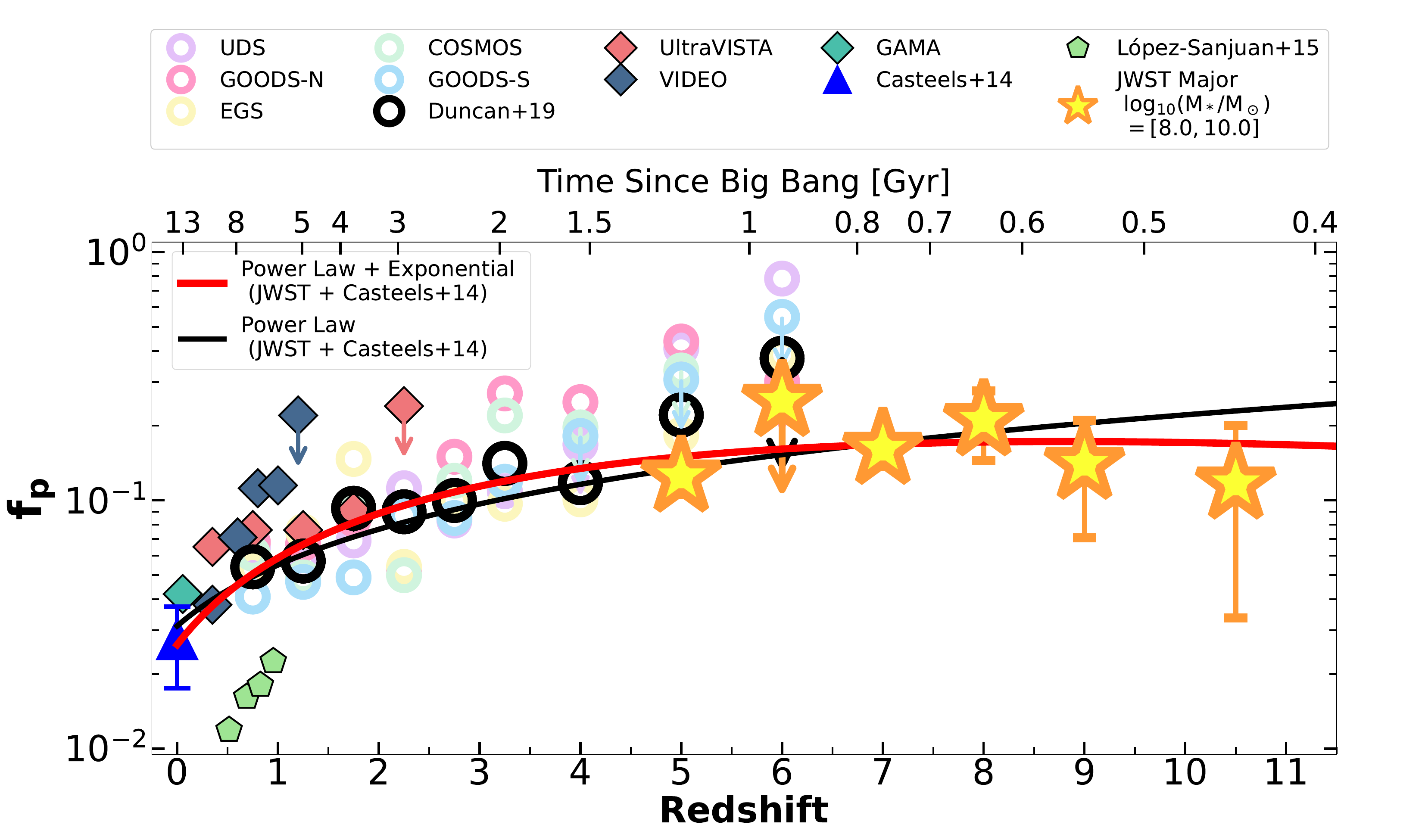}
  \caption{A compilation of measured pair fractions as a function of redshift in the literature, from \citet{Duan2025}. New high-redshift major merger fractions ($\rm f_p$) from the JWST analysis of \citet{Duan2025} are shown using the yellow-orange stars. The stellar mass range of this analysis is 10$^{8}$ M$_{\odot}$ $<$ M$_{\star}$ $<$ 10$^{10}$ M$_{\odot}$. Green pentagons show data from \citet{Lopezsanjuan2015} for galaxies with $M_{\rm B}$ $<$ -20. Rings represent data from \citet{Duncan2019}, with the ring colours indicating different fields and the black rings representing the combined results for stellar masses greater than 10$^{10.3}$ M$_{\odot}$. Diamonds show data from \citet{Conselice2022} for stellar masses greater than 10$^{11}$ M$_{\odot}$. Blue triangles show results from \citet{Casteels2014}. All literature values except for the results from \citet{Conselice2022} show derived $\rm f_p$ measurements corresponding to major mergers, while \citet{Conselice2022} presents $\rm f_p$ measurements from major and minor mergers. The black and red lines show fits to the JWST measurements of \citet{Duan2025} using a power-law (equation 19 in \citet{Duan2025}) and a variant which uses a power-law plus an exponential (equation 20 in \citet{Duan2025}), with the \citet{Casteels2014} data used as a zero point. Credit: Figure 4 in Qiao Duan et al. `Galaxy Mergers in the Epoch of Reionization I: A JWST Study of Pair Fractions, Merger Rates, and Stellar Mass Accretion Rates at $z$ = 4.5 - 11.5'. Monthly Notices of the Royal Astronomical Society (2025) doi:10.1093/mnras/staf638 $\copyright$ Qiao Duan et al. This article is distributed under the terms of a \href{https://creativecommons.org/licenses/by/4.0/}{CC BY 4.0 license} and the original article can be accessed at \protect\url{https://doi.org/10.1093/mnras/staf638}.}
  \label{fig:mergerfractions}
\end{figure}

Figure \ref{fig:mergerfractions} shows a compilation from the literature of the merger fraction across cosmic time. While past studies have indicated that the major merger fraction increases out to at least $z\sim2-3$, the behaviour at very high redshift remains unclear. While some studies, based on recent observations from the James Webb Space Telescope (JWST), suggests that this upward trend continues out to $z\sim6$, followed by a statistically flat evolution until $z\sim11$ \citep[see e.g. data in Figure \ref{fig:mergerfractions} from][]{Duan2025}, other studies suggest that the merger fraction may not evolve towards high redshift \citep[e.g.][]{Dalmasso2024}. The discrepancies between these studies is likely to be driven by the methods (e.g. close pairs vs morphological parameters) employed to detect mergers. Finally, in contrast with the major merger fraction in massive galaxies, the observed minor merger fraction appears to be consistent with no redshift evolution \citep[e.g.][]{Lopezsanjuan2012,Ventou2019}. 

We complete this section by considering how these empirical results compare to current theoretical models since, in the context of the hierarchical paradigm, the observed merger fraction is a potential constraint on the efficacy of galaxy formation models. As noted above, the merger tree of any given object within a simulation can be defined with relative ease. However, even when simulations are compared to observations in a consistent manner, via, for example, forward modelling of synthetic images, observed and theoretical merger fractions do not always agree \citep[e.g.][]{Wang2020}. Agreement tends to be better at the high mass end ($M_{\star} > 10^{11}$ M$_{\odot}$) at $z\lesssim 2$ (where galaxies are best resolved in simulations), with significant discrepancies at lower stellar masses in this redshift range \citep[e.g.][]{Bertone2009} and all stellar masses at higher redshifts \citep[e.g.][]{Duan2025}. 

The disagreements are likely to be driven by a combination of factors. In simulations, key galaxy properties such as stellar mass are strongly influenced by the implementation of baryonic processes, such as feedback from supernovae and active galactic nuclei \citep[e.g.][]{Kaviraj2017}. This, in turn, can result in significantly different predictions for the merger fractions, which can vary by up to factors of a few \citep[e.g.][]{Bertone2009}. Several other reasons could contribute to these discrepancies, albeit not as significantly as the impact of baryonic feedback. For example, cosmological simulations do not offer the same volumes as large surveys, which could induce uncertainties in the merger fractions at high stellar masses. In a similar vein, the use of photometric (rather than spectroscopic) redshifts is likely to induce uncertainties in the redshift evolution of the merger fraction because precise distances to the galaxies in question are not known \citep[e.g.][]{Lopezsanjuan2015}. The discrepancies between theoretical and observed merger fractions are likely to be driven more by such methodological issues rather than a fundamental problem in the standard structure-formation paradigm.  



\section{The impact of mergers on galaxy evolution}

We now consider how the presence of mergers may alter key aspects of a galaxy's evolution. We begin by considering how the merger history of a galaxy may alter its morphology over its lifetime. We then explore the impact that mergers have on galaxy SFRs, the accretion rates onto central supermassive black holes and the cumulative impact of these processes in terms of the stellar and black-hole mass that exists in the Universe today. 


\subsection{Morphological transformation}
\label{sec:morphological_transformation}

Observational work demonstrates that the morphological mix of massive galaxies evolves over cosmic time. While the massive-galaxy population at intermediate and high redshift ($z>1$) is largely composed of rotationally-supported \citep[e.g.][]{Buitrago2014}, disc-dominated late-type systems \citep[e.g.][]{Vanderwel2011}, their counterparts in the nearby Universe are more dispersion-dominated and, particularly at the highest stellar masses, show a preponderance of systems with early-type morphology \citep[e.g.][]{Buitrago2013,Lintott2011}. The current consensus, driven mainly by theoretical work in the context of hierarchical structure formation, suggests that this morphological transformation is likely to be driven largely by mergers. 

\begin{figure}
  \centering
  \includegraphics[width=\textwidth]{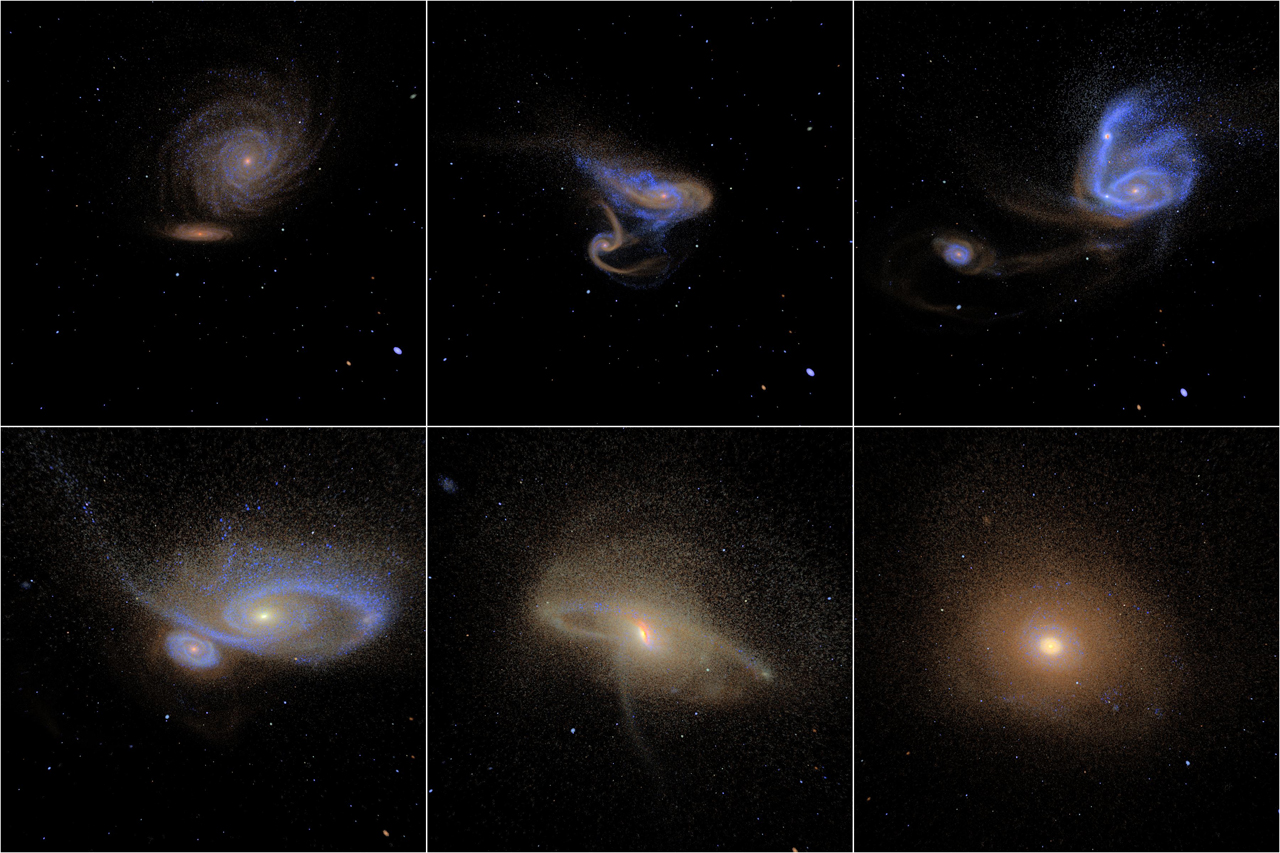}
  \caption{Images of the time sequence (clockwise from the top left) of a simulation of a merger between two disc galaxies. The remnant is an early-type galaxy (bottom right hand panel). The time sequence spans a timescale of around 1 Gyr. Credit: P. Jonsson (Harvard-Smithsonian Center for Astrophysics), G. Novak (Princeton University), and T. J. Cox (Carnegie Observatories, Pasadena, California).}
  \label{fig:timesequence}
\end{figure}

Early N-body simulations \citep[e.g.][]{Toomre1972,White1978,Negroponte1983} showed that mergers of discs (with similar stellar mass), especially those that are embedded in extended dark matter halos \citep[e.g.][]{Barnes1988}, produce remnants whose properties resemble those of nearby early-type galaxies. Dynamical friction during the early stages of a merger acts to bring the merging galaxies closer together. During the later stages of this process, the gravitational potential varies rapidly compared to the typical dynamical timescales of the stellar orbits, which acts to make the orbits in the remnant more chaotic on short timescales. The principal consequences of this process are to reduce or wash out pre-existing  structure in the progenitors (e.g. spiral arms) and move material towards the central regions of the remnant. The end result is a system which has a central mass (and therefore light) concentration and which is largely devoid of fine structure, akin to what is seen in early-type galaxies in the nearby Universe. Furthermore, as described in Section \ref{sec:visual_inspection}, mergers produce tidal features (e.g. tidal tails, streams and shells) which indicate the presence of such events in the recent dynamical history of the galaxy in question. Figure \ref{fig:timesequence} shows several snapshots from a time sequence of a simulated merger between two disc galaxies which produces an early-type remnant.  


The postulate that mergers contribute to the morphological transformation seen in massive galaxies and create early-type systems is supported by observations. A direct observational indicator of merger-driven morphological transformation is the presence of tidal features around early-type galaxies. In deep optical imaging, the vast majority (at least 70 per cent) of massive early-types in the nearby Universe show tidal tails and streams, shells and internal asymmetries indicating that most (if not all) of these galaxies have experienced recent merger events \citep[e.g.][]{vandokkum2005,Tal2009,Duc2015}. This is supported by other indirect evidence of the importance of merging in driving morphological transformation. For example, the build up of the luminosity density of massive red galaxies (which are dominated by early-type systems) is difficult to explain without invoking merger activity \citep[e.g.][]{Baldry2004,Faber2007}. Similarly, the fraction of massive early-types appears to increase in denser environments, where interactions are likely to be more frequent \citep[e.g.][]{Dressler1980,Bamford2009}, with merging thought to be a key contributor to the creation of this correlation \citep[e.g.][]{Pfeffer2023}. Observations therefore provide strong qualitative support for the notion that mergers are important drivers of morphological transformation in massive galaxies over cosmic time. 

However, it is interesting to note that, notwithstanding their simple appearance, massive early-type galaxies in the local Universe show a large diversity in their stellar kinematics. A detailed analysis, using integral field spectroscopy, of a volume-limited sample of massive early-types in the local Universe, reveals that this population can be split into two distinct classes of `slow' and `fast' rotators \citep[e.g.][]{Cappellari2011}. Note that, while the fast rotators do not have spiral arms or discs (i.e. they are indeed morphologically early-type), some of them can exhibit global rotation that is as fast as what is seen in disc galaxies. Around 86 per cent of the massive early-type population are classified as fast rotators. 

High-resolution `zoom-in' simulations of individual massive galaxy halos \citep{Naab2014} are able to produce good agreement with the kinematical diversity demonstrated in \citet{Cappellari2011}. They show that the final kinematic properties of the massive early-type galaxies that exist in these halos at the present day are sensitive primarily to the fraction of stars that has formed in-situ from accreted gas and to the properties and look-back times of major mergers. There appear to be three principal channels for forming fast rotators: (a) forming a significant ($>$40 per cent) fraction of star in-situ and having late minor mergers but no major merger since $z\sim1$, (b) experiencing a late gas-rich major merger, which spins up the merger remnant (or leaves an existing highly-spinning galaxy unchanged) and (c) experiencing one or more gas-poor major mergers which spin up the merger remnant (or leave an existing highly-spinning galaxy unchanged). In a similar vein, there are also three primary pathways to forming a slow rotator: (a) experiencing gas-rich major mergers, the net effect of which is to spin down the remnant (b) undergoing at least one recent major and multiple minor mergers, which together spin the remnant down and (c) experiencing multiple minor mergers but no major merger at $z<2$. 

\begin{figure}
  \centering
  \includegraphics[width=\textwidth]{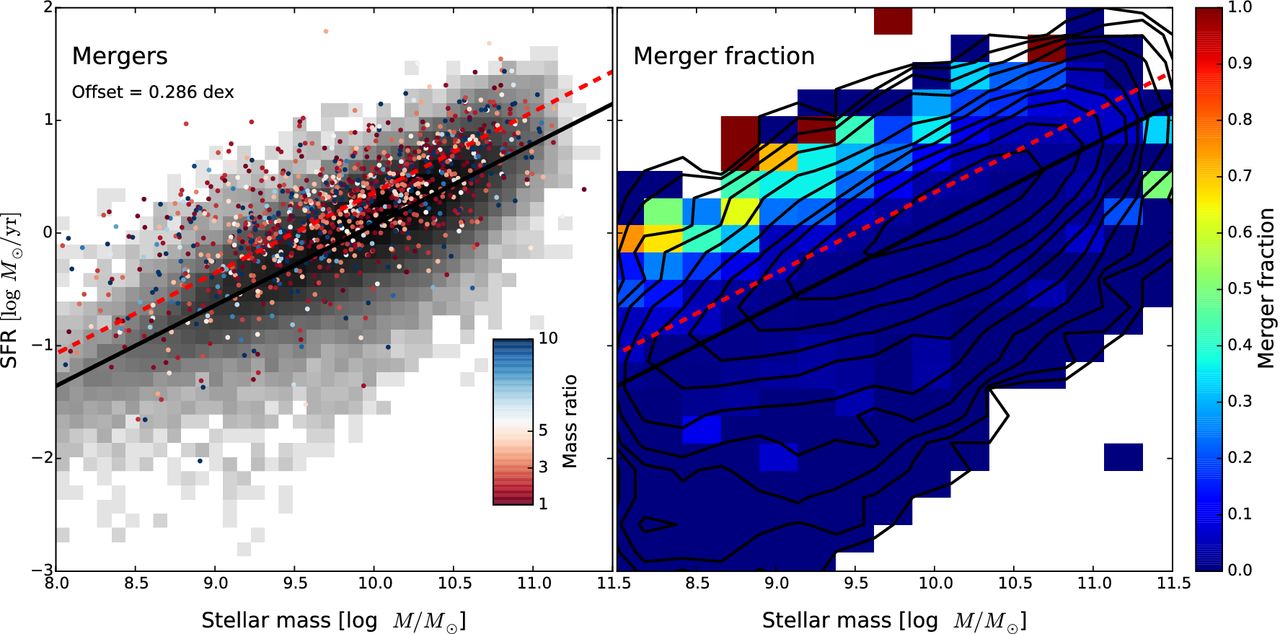}
  \caption{SFR as a function of $M_{\star}$ in the nearby Universe, from \citet{Willett2015}. \textbf{Left:} The grey-scale heatmap shows the entire galaxy population. The coloured points show 2978 merging galaxies from \citet{Darg2010}. Mergers are colour-coded by the mass ratio of the primary and secondary galaxies (see legend). The merging population (red dashed line) has a higher SFR by $\sim$0.3 dex compared to all star-forming galaxies (black solid line). \textbf{Right:} Star-forming galaxies binned and colour-coded by merger fraction (i.e. the number of mergers divided by the number of star forming galaxies). Overplotted lines are the same as in the left hand panel. Of the galaxies that lie furthest above the star formation main sequence, more than 50 per cent can be in merging systems. Credit: Figure 6 in Kyle W. Willett et al. `Galaxy Zoo: the dependence of the star formation–stellar mass relation on spiral disc morphology.' Monthly Notices of the Royal Astronomical Society (449) 1 (2015): 820–827.}
  \label{fig:sfms_mergers}
\end{figure}

The theoretical work we have described so far has focused on individual galaxies that are merger remnants, created either via idealised binary simulations of mergers or zoom-in simulations of individual massive galaxies selected from cosmological volumes. Such scenarios consolidate the notion that the morphological changes in at least a representative population of mock galaxies at the present day is strongly connected to merging. However, it is desirable to construct an aggregate view of how morphological change proceeds in the general galaxy population as a function of cosmic time, and how that is driven by different types of mergers. 

Using the Horizon-AGN cosmological hydro-dynamical simulation, \citet{Martin2021} quantitatively track the transformation of morphology in massive galaxies over the lifetime of the Universe. Morphology is parameterised by $V/\sigma$, the ratio of the mean rotational velocity ($V$) and the mean velocity dispersion ($\sigma$). Lower values of $V/\sigma$ correspond to systems that are more dispersion-dominated i.e. those that have more early-type morphology, while higher values of $V/\sigma$ correspond to systems that are more rotationally supported i.e. have more late-type morphologies. Measuring changes in $V/\sigma$ then allows the process of morphological transformation to be studied in a global sense and connected to the properties of mergers over cosmic time. 

A key conclusion in \citet{Martin2021} is that essentially all the morphological evolution in massive galaxies that end up with early-type morphology at the present day is driven by mergers with mass ratios greater than 1:10. In agreement with the simulations of \citet{Naab2014}, they find that major mergers alone cannot create today's early-type population. Minor mergers, with mass ratios between 1:10 and 1:4 drive around a third of all morphological transformation over cosmic time and become its dominant driver after $z\sim1$. Interestingly, prograde mergers, where the orbit of the secondary companion is in the same direction as the spin of the primary, produce milder morphological changes than retrograde mergers. For example, the average morphological change due to retrograde mergers is around twice that due to their prograde counterparts at $z\sim0$. The morphology of remnants depends on the gas fraction in a merger, with the remnants of gas-rich mergers being able to re-grow discs from gas left over in the merger. Finally, galaxies of a given stellar mass, regardless of whether they are late or early-types, actually experience similar merger histories. The survival of massive discs to the present day is driven by acquisition of cold gas (either via gas-rich mergers or cosmological accretion) and a preponderance of prograde mergers in their merger histories.

\subsection{Impact on star formation activity and black-hole growth}

The gravitational torques created in mergers cause gas to lose angular momentum, driving it towards the central regions of the remnant where it can trigger starbursts \citep[e.g.][]{Cox2008,DiMatteo2007}. The general theoretical expectation is, therefore, that the SFR is likely to be enhanced during the merger process (compared to what it typically is when galaxies are not merging). At low redshift, observational work shows that, while high SFRs are often associated with mergers, not all galaxies that are merging exhibit significant star formation activity. The right hand panel of Figure \ref{fig:sfms_mergers} shows that the fraction of galaxies that reside in merging systems increases with distance from the SFR vs $M_{\star}$ relation (often referred to as the `star formation main sequence'). Indeed, in galaxies that lie significantly above the star formation main sequence, more than 50 per cent tend to be mergers \citep[e.g.][]{Willett2015}. 

\begin{figure*}
  \centering
  \includegraphics[width=0.45\textwidth]{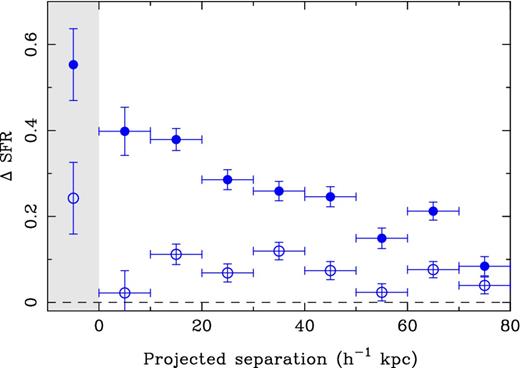}
  \includegraphics[width=0.45\textwidth]{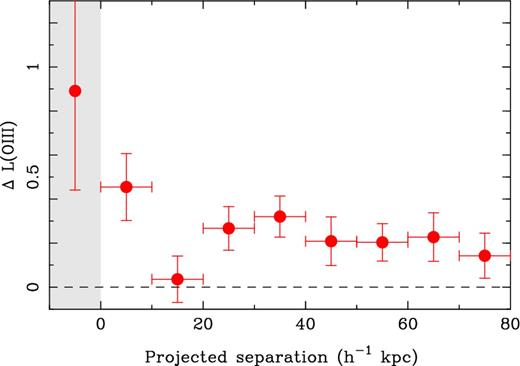}
  \caption{\textbf{Left:} The enhancement in the SFR in close pairs of star-forming galaxies, from \citet{Ellison2013}. Filled circles show the enhancements measured in the SDSS fibre (which samples a 3 arcsecond region in the centre of the galaxy), and open circles show the enhancements outside the fibre. The points in the grey shaded box show the enhancements for galaxies in post-merger systems. \textbf{Right:} The enhancement in the [O III] luminosity in close pairs of star-forming galaxies, from \citet{Ellison2013}. AGN are classified using optical emission line ratios following the criteria in \citet{Kewley2001}. Credit: The left and right hand panels are taken from Figures 6 and 11 in Sara L. Ellison et al. `Galaxy pairs in the Sloan Digital Sky Survey – VIII. The observational properties of post-merger galaxies.’ Monthly Notices of the Royal Astronomical Society (435) 4 (2013): 3627-3638. Reproduced with permission from the Oxford University Press on behalf of the Royal Astronomical Society.}
  \label{fig:sfenhancement_mergers}
\end{figure*}

Not unexpectedly, the enhancement of star formation activity during ongoing mergers increases as the separation between the merging galaxies decreases \citep[e.g.][]{Patton2011,Scudder2012}. Figure \ref{fig:sfenhancement_mergers} shows that, while SFR enhancements of $\sim$ 25 per cent already exist at large projected separations ($\sim$80 kpc), these rise to around a factor of 2.5 for separations less than 20 kpc and up to a factor of 3.5 in post-mergers \citep{Ellison2013}. While both major and minor mergers can produce significant enhancements in the SFR, the enhancements are typically larger and extend to greater separations in major mergers \citep[e.g.][]{Ellison2008}.  

\begin{figure}
  \centering
  \includegraphics[width=0.6\textwidth]{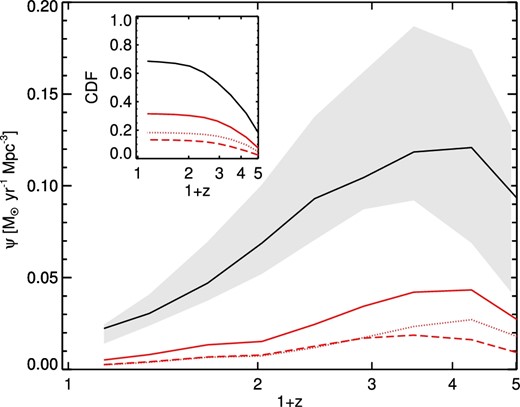}
  \caption{The cosmic SFR density from the Horizon-AGN simulation (black line) and the contribution due to major and minor mergers (solid red line), just major mergers (dashed red line) and just minor mergers (dotted line), from \citet{Martin2017}. The inset shows the cumulative fraction of stellar mass formed due to major and minor mergers (solid red line), just major mergers (dashed red line), just minor mergers (dotted red line) and other processes (solid black line). The grey filled area indicates the 3$\sigma$ confidence region from observations \citep[][]{Hopkins2006}. Credit: Figure 3 in Garreth Martin et al. `The limited role of galaxy mergers in driving stellar mass growth over cosmic time.' Monthly Notices of the Royal Astronomical Society: Letters (472) 1 (2017): L50–L54.}
  \label{fig:sfhistory_mergers}
\end{figure}

The situation at higher redshifts is different from what is seen in the nearby Universe. At $z\sim2$, which is close to the epoch of peak cosmic star formation \citep[e.g.][]{Madau2014}, the enhancement of star formation in mergers is much lower than what is observed at low redshift. At this epoch, only a few per cent of galaxies that lie above the star formation main sequence are merger-driven starbursts. Furthermore, around 10 per cent or less of the overall SFR density in massive galaxies can be ascribed to the merger process \citep[e.g.][]{Rodighiero2011,Lofthouse2017}, several factors lower than what is found at low redshift \citep[e.g.][]{Kaviraj2014b}. 

The differences at low and high redshift are driven by the differing structures of discs at these epochs. Massive high-redshift discs have significantly larger gas fractions  \citep[e.g.][]{Daddi2010,Tacconi2010} which makes them more prone to disc instabilities that can induce strong nuclear inflows and the formation of clumps \citep[e.g.][]{Bournaud2010}.  At low redshift, discs have mild central mass inflows which can increase by more than an order of magnitude during mergers. In contrast, high-redshift discs have stronger instability-driven inflows from the outset, as a result of which the enhancement of the inflow rate induced by mergers is much weaker \citep[e.g.][]{Fensch2017}. Furthermore, while mergers increase the compressive turbulence which aids in triggering central starbursts, high-redshift clumpy discs are already turbulent, as a result of which mergers cause only moderate increases in the turbulence of the disc \citep[e.g.][]{Bournaud2010,Hopkins2013,Fensch2017}. Together, these factors cause the SFR enhancements due to mergers to be much milder at high redshifts than in the nearby Universe. 

Given that a significant fraction of the stellar mass that exists in the local Universe forms around the epoch of peak cosmic star formation, the inability of mergers to enhance star formation activity around these redshifts means that most of the stars that exist at the present day are not formed as a result of mergers (but rather simply through gas accretion). In this context it is desirable to quantify what fraction of the stellar mass at the present day is attributable to major and minor mergers. \citet{Martin2017} have used the Horizon-AGN cosmological simulation to study the contribution of major and minor mergers to the star formation history of massive galaxies over the lifetime of the Universe. Figure \ref{fig:sfhistory_mergers}, from \citet{Martin2017}, shows the cosmic SFR density as a function of redshift (solid black line), with the contributions of all mergers (solid red line), major mergers (red dashed line) and minor mergers (red dotted line). The inset shows the cumulative fraction of stellar mass formed due to major and minor mergers (solid red line), just major mergers (dashed red line), just minor mergers (dotted red line) and other processes (solid black line). This indicates that only around 30 per cent of the stellar mass in today's massive galaxies forms as a direct result of major and minor mergers, with the rest forming via gas accretion. 

We complete our discussion by considering how mergers affect the growth of the supermassive black holes that reside in the central regions of galaxies. As described above, the theoretical expectation is that mergers drive material towards the central regions of the remnant, although this process becomes weaker in systems with high gas fractions which dominate the early Universe. Once the gas loses enough angular momentum, it can be accreted on to the black hole. As some of the gravitational potential energy from the accreted material is released this process creates an `active galactic nucleus' (AGN) and fuels an episode of emission from the black hole \citep[e.g.][]{Johansson2009,Byrnemamahit2023}. 

From an observational point of view, there is some evidence that, in nearby galaxies, the black hole accretion rates are enhanced, although the trends tend to be fairly weak \citep[e.g.][]{Ellison2013,Kaviraj2014b}. For example, the right-hand panel of Figure \ref{fig:sfenhancement_mergers} shows that the black hole accretion rate, traced via the luminosity of the [O III] emission line, shows an increase with decreasing projected distance in close pairs. The trend is similar to what is found in the SFR (see the left-hand panel of this figure) although much milder. However, when the current literature is considered in totality, there is no clear consensus about the connection between AGN activity and mergers. This is most likely driven by the heterogeneity of the galaxy samples used and the methods employed to identify AGN. \citet{Villforth2023} have combined the results of 33 observational studies from the literature to determine the relationship between mergers and AGN activity. These studies span a large redshift range ($0.25<z<2$) and bolometric AGN luminosities (41.5 $<$ log (L$_{\rm bol}$ / erg s$^{-1}$) $<$ 48). The broad conclusions of this study are that the contribution of mergers to the AGN population appears significant only in subsets of AGN types (mainly radio AGN and AGN that reside in red galaxies). There also does not appear to be a correlation between merger fraction and AGN luminosity. This aggregate observational view appears to suggest that the overall role of mergers in driving black hole accretion may be a minor one. 

In this context two observational results are worth noting. First, the masses of central supermassive black holes appear to be tightly correlated with the stellar masses of their host galaxies \citep[e.g.][]{Haring2004}. Second, the shape of the cosmic black hole accretion rate density closely tracks that of the cosmic SFR density \citep[e.g.][]{Madau2014}. Taken together, this indicates that the growth of the stellar mass of galaxies moves in lockstep with the growth of their supermassive black holes. Given that both star formation and black hole growth are triggered by the same process, i.e. gas inflows, it is reasonable to suggest that the trends seen in the SFR enhancement may be mirrored to some extent in the enhancement of the black-hole accretion rate. Recall, from the discussion above that the overall stellar mass growth in the Universe is not driven by merging which suggests that the same may be true of black hole growth, consistent with the conclusions of \citet{Villforth2023}. 

\begin{figure*}
  \centering
  \includegraphics[width=0.45\textwidth]{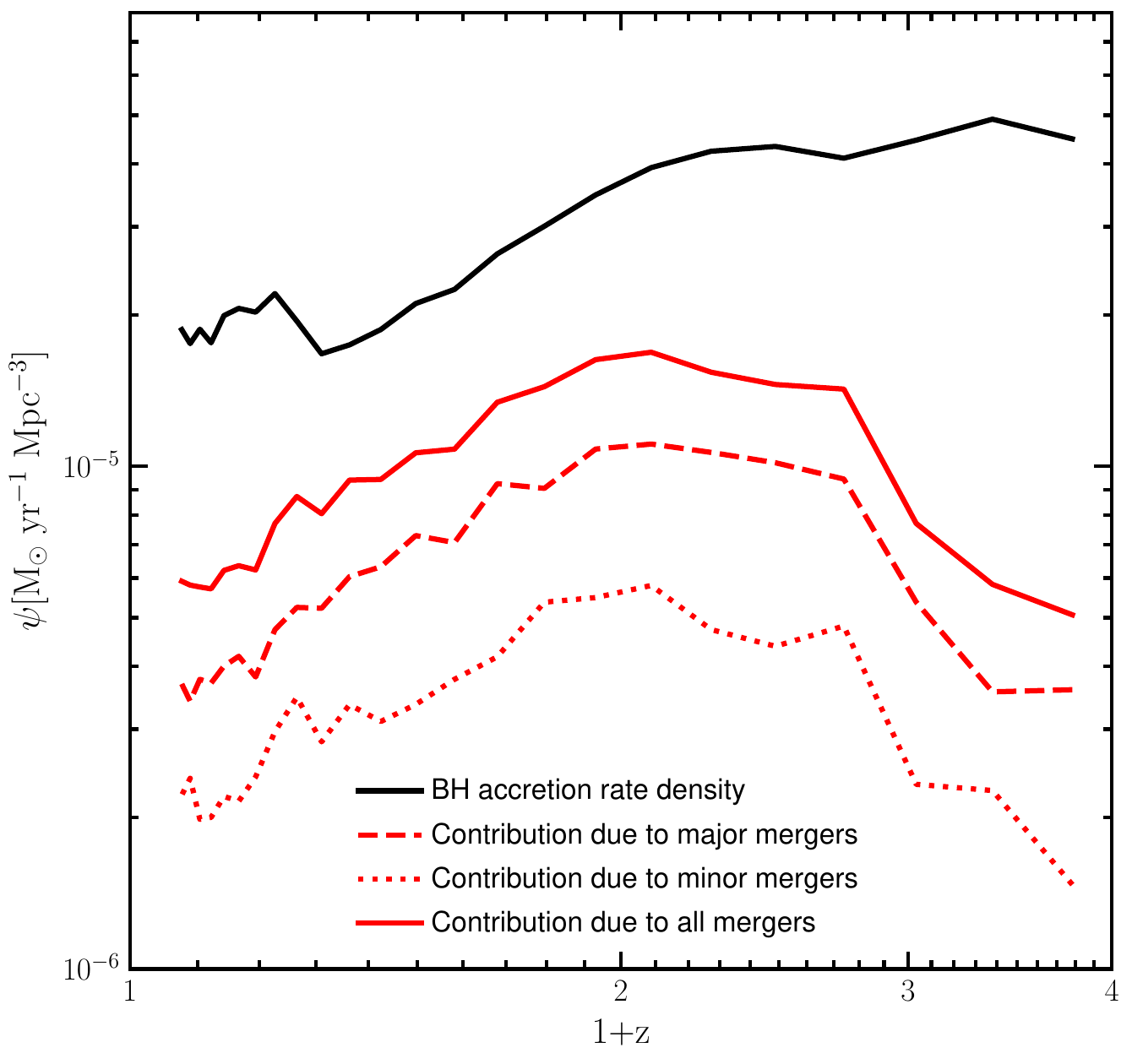}
  \includegraphics[width=0.44\textwidth]{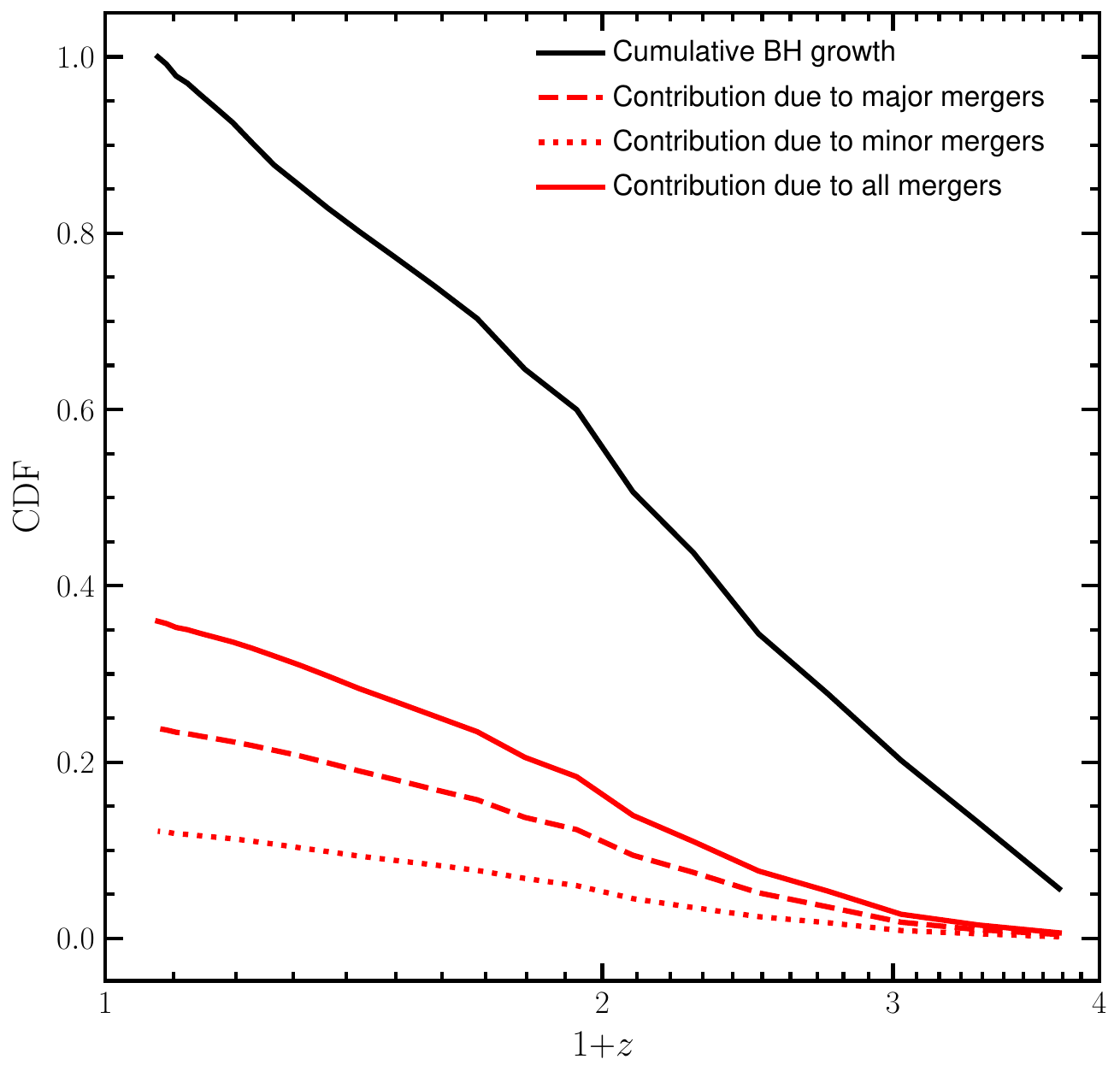}
  \caption{\textbf{Left:} The black-hole accretion rate density in massive galaxies as a function of redshift from the Horizon-AGN simulation (black), from \citet{Martin2018_BH}. The red lines indicate the fraction that is a direct result of major (dashed line), minor (dotted line), and major + minor (i.e. all) mergers (solid line). The small jump in accretion at low redshift corresponds to the introduction of an additional grid refinement level at $z = 0.26$. \textbf{Right:} The cumulative fraction of black-hole mass in today's massive galaxies that has already been assembled as a function of redshift (black line), from \citet{Martin2018_BH}. The contributions from major mergers, minor mergers, and major + minor (i.e. all) mergers are shown using the dashed, dotted, and solid red lines, respectively. Only $\sim$35 per cent of the black-hole mass in massive galaxies at the present day is directly attributable to merger activity. $\sim$22 per cent is driven by major mergers and $\sim$13 per cent is driven by minor mergers. Thus, the bulk ($\sim$65 per cent) of the black-hole mass build-up over cosmic time is unrelated to merging. Credit: The left and right-hand panels are adapted from Figures 8 and 9 in Garreth Martin et al. `Normal black holes in bulge-less galaxies: the largely quiescent, merger-free growth of black holes over cosmic time'. Monthly Notices of the Royal Astronomical Society (476) 2 (2018): 2801–2812.}
  \label{fig:bharhistory_mergers}
\end{figure*}

\citet{Martin2018_BH} have used the Horizon-AGN cosmological simulation to track the effects of major and minor mergers on the growth of black holes in massive galaxies over cosmic time. Their findings are summarised in Figure \ref{fig:bharhistory_mergers}. The left-hand panel shows the black-hole accretion rate history of massive galaxies (solid black line) and the contributions due to major and minor mergers (solid red line), just major mergers (dashed red line), just minor mergers (dotted red line). This suggests, in agreement with the results of other similar studies \citep[e.g.][]{McAlpine2018}, that mergers are responsible for only a minority of the growth on to black holes at any epoch. The right-hand panel shows the cumulative version of this evolution and suggests that, when major and minor mergers are considered together, around 35 per cent of the black hole mass in today's massive galaxies is directly attributable to merger activity.





\section{Summary and outlook}

In the hierarchical paradigm of structure formation, the evolution of the observable Universe is significantly influenced by the attractive force of gravity. A fundamental aspect of this evolution, therefore, is frequent merging between galaxies. 
Observational techniques to find mergers include identifying galaxy pairs in close proximity to each other, visual inspection of galaxy images to identify signatures of mergers (e.g. tidal features, shells and internal asymmetries) and using morphological parameters (e.g. Asymmetry and the Gini coefficient) which are sensitive to parts of the merger process. 

Observational work using large-scale surveys indicates that the merger fraction increases with redshift, potentially out to $z\sim6$, followed by flat evolution out to $z\sim11$. While the merger histories of mock galaxies in simulations are known, comparisons of the merger fractions between observations and theory are often in tension with each other. This is because the predicted merger fractions can be sensitive to differing treatments of baryons in different simulations and also due to the limitations of both simulations and data (e.g. simulation volumes being much smaller than those traced by large surveys and the use of photometric redshifts in deriving empirical merger fractions).

The most significant impact of mergers is in transforming galaxy morphology. Over time the morphology of massive galaxies shows a gradual shift from being rotation-dominated (i.e. late-type) at high redshift to more dispersion-dominated (i.e. early-type) at low redshift (with early-type galaxies being the dominant morphological class at the highest stellar masses in the local Universe). Virtually all the morphological evolution in massive galaxies that end up with early-type morphology at the present day is due to mergers with mass ratios greater than 1:10. Major mergers alone do not create today's early-type population. Minor mergers drive around a third of all morphological transformation over cosmic time and become its dominant driver after $z\sim1$. 

Simulations show that the merger process causes material to lose angular momentum and fall towards the central regions of the remnant. This can trigger enhancements in both star formation activity and black-hole accretion, above what would have happened had the merger not taken place. However, both observational and theoretical work shows that, over cosmic time, only around a third of the stellar and black-hole mass seen in today's massive galaxies formed as a direct result of mergers. 

While much is understood about the role of mergers in massive galaxies, less is known about how merging impacts dwarf galaxies over the lifetime of the Universe. In this context, forthcoming deep-wide surveys, such as the Legacy Survey of Space and Time \citep[LSST;][]{Ivezic2019} and Euclid \citep{Gardner2023}, are poised to revolutionise our understanding of dwarf galaxy evolution by providing unbiased statistical samples of dwarfs outside our local neighbourhood. Given the unprecedented (petabyte-scale) data volumes expected from such surveys, many of the techniques that have traditionally been used to identify mergers (e.g. visual inspection) may become prohibitively time-consuming to deploy on such datasets. Machine-learning techniques \citep[e.g.][]{Pearson2019,Martin2020,Uzeirbegovic2020} are likely to become important for dwarf-galaxy studies that are underpinned by these new datasets.  

Some recent work offers a preview of what studies of mergers and interactions may find in the dwarf regime. For example, \citet{Lazar2024} have studied around 250 nearby ($z<0.06$) dwarfs, using survey data from the Hyper Suprime-Cam \citep{Aihara2018a} which reaches similar depths as LSST over 2 deg$^2$. They show that while eqs. \ref{eq:as_merger} and \ref{eq:gm20_merger} are able to find merging systems in massive galaxies, some modifications are required in the dwarf regime. In dwarf early-types which have $A>0.05$, more than 50 per cent of systems show signs of interactions. The corresponding threshold for dwarf late-types is $A>0.08$. However, combinations of Gini and $M_{\rm 20}$ are unable to separate interacting and non-interacting dwarfs, unlike in the massive galaxy regime. The lower asymmetry thresholds required to select interacting dwarfs is likely to be driven by the fact that their shallower potential wells makes it easier to induce structural changes in dwarfs than in their massive counterparts. The study of dwarf galaxies using deep-wide surveys will significantly improve the state-of-the-art, in terms of our understanding of how mergers drive galaxy evolution, in the coming years.

\begin{ack}[Acknowledgments]
 TThe author acknowledges support from the STFC (grant numbers ST/Y001257/1 and ST/X001318/1) and a Senior Research Fellowship from Worcester College Oxford.  
\end{ack}


\bibliographystyle{Harvard}
\bibliography{references}

\end{document}